# Extracting Signals of Higgs Boson From Background Noise Using Deep Neural Networks

Muhammad Abbas[a], Asifullah Khan[b,c,*], Aqsa Saeed Qureshi[b], Muhammad Waleed Khan[b,c]

[a]Department of Physics and Applied Mathematics,
[b]Department of Computer and Information Sciences,
[c]PIEAS Artificial Intelligence Center (PAIC),
Pakistan Institute of Engineering and Applied Sciences, Islamabad, Pakistan

**Abstract**

Higgs boson is a fundamental particle, and the classification of Higgs signals is a well-known problem in high energy physics. The identification of the Higgs signal is a challenging task because its signal has a resemblance to the background signals. This study proposes a Higgs signal classification using a novel combination of random forest, auto encoder and deep auto encoder to build a robust and generalized Higgs boson prediction system to discriminate the Higgs signal from the background noise. The proposed ensemble technique is based on achieving diversity in the decision space, and the results show good discrimination power on the private leaderboard; achieving an area under the Receiver Operating Characteristic curve of 0.9 and an Approximate Median Significance score of 3.429.

## 1. Introduction

Higgs boson, in the standard model of particle physics, is an elementary particle. Higgs boson is a quantum excitation of the Higgs field. The discovery of Higgs boson is considered as a remarkable milestone in the history of particle physics. All the elementary particles interact with the Higgs field via a Higgs mechanism. Moreover, mass is not an intrinsic property of these subatomic particles; they get it via Higgs mechanism [1]. One of the most important models in this context is MSSM [2]. Higgs particle may be a gateway to many predictions for beyond standard model explorations.

∗Corresponding author.
*E-mail address:* asif@pieas.edu.pk

In May 2014, Higgs Boson Machine Learning Challenge, a competition hosted by Kaggle and proposed by the ATLAS collaboration was held. The competitors were required to build an algorithm that could classify Higgs signals from those of background signals [3]. The dataset provided for the competition consisted of 30 features, out of which 17 were fundamental features obtained via real measurements of the detector, whereas 13 features were derived via combining primary features by applying some mathematical formulae [3].

Higgs Boson Machine Learning Challenge was a "learning to discover" task, classifying the events of interest (signals), from those of already predicted by the experiment in the past (background signals). The training data was generated through simulator implementing theoretical predictions. Search of Higgs signals from background is very complex because the Large Hadron Collider (LHC) generates a huge volume (petabytes) of raw data every year. A comprehensive study of Higgs boson search is given in [4]. Furthermore, Higgs indication signals are very rare in background signals. The competition was led for the purpose to optimize the Approximate Median Significance (AMS) metric as an evaluation measure. The challenge helped in making both qualitative as well as quantitative advances over optimization of the AMS.

*1.1. Higgs Boson*

The discovery of Higgs boson was made by CMS and ATLAS collaborations at CERN in July 2012. CMS and ATLAS experiments are performed by two separate communities of physics [5]. The evidence comes from its three decay modes $H \rightarrow zz^* \rightarrow 4l$, $H \rightarrow \gamma\gamma$, $H \rightarrow ww^*$ [6]. The mass of Higgs is $M_H$=125 GeV. Because of the known mass, the properties of Higgs boson are uniquely determined, and all Higgs couplings are fixed. The predicted decay branching fractions of Higgs boson depends on its interaction strength with its decay products [7]. There are different decay modes available for Higgs boson. For Higgs of mass 125 GeV, the branching fraction for $H \rightarrow b\bar{b}$ reach up to 56.9% and for $H \rightarrow \tau\bar{\tau}$ to 6.2%.

## 1.2. Higgs Boson Searches at LHC

The production of Higgs boson at Large Hadron Collider (LHC) involves heavy particles, like the W & Z bosons and heavy fermions. The Higgs boson production at LHC is a very rare process because of its small production cross section. Before the Higgs boson detection, some precise calculations for the cross section of Higgs boson were performed [8]. The searches for Higgs boson lead to severe backgrounds.

The production of Higgs boson through gluon fusion is significant due to the large abundance of gluons at LHC. Higgs may yield from the radiation of W or Z bosons from incoming quarks which then fuse to yield a Higgs particle. This process is also called vector boson fusion (VBF). Taken from ATLAS experiment, in which the Higgs boson decays to four lepton channel, and using the Run1 data, the Higgs invariant mass peak is near 125 GeV [9], whereas the background mass peak from Z boson is at 91 GeV.

## 1.3. Data Details

The data set can be represented as $D = \{x_i, y_i\}_{i=1}^n$ ($x_i = R^n$, $y_i \in \{0, 1\}$). Suppose $n$ quantities are measured from an event $X = (x_1, ..., x_n)$, where $n$ is the number of features recorded for a single event ($n=30$ in this case). The joint probability of vector $X$ is dictated by the type of event, say, $F(X|b)$ for background events and $F(X|s)$ for signal event [7].

The dataset is acquired from the kaggle website. The kaggle is the home of competitions for machine learners and data scientists. The dataset consists of 250000 and 550000 of training and testing samples respectively [3].

There were total of 30 features available in the original data set. In dataset, all the missing values that were invalid for the event, were replaced with -999.0. In proposed work, these missing values are replaced with "zero" and "mean" to analyze its effect on classification.

The competition was led for the purpose to optimize the AMS score. Any of the competitors training and optimizing the model would then submit the classifier. The only output would be the AMS score for every submitted classifier [10]. AMS has often been used as a measure of statistical significance and experimental sensitivity. $AMS = \sqrt{2(s+b)ln(1 + \frac{s}{b}) - s}$, where $s$ and $b$ are true positive and false positive rates, respectively. For $s << b$, it is reduced to $s/\sqrt{b}$.

*1.4. Related Work*

Several approaches had been used to optimize the AMS. The AMS score of the top ten participant's ranges from 3.76 to 3.80. Most of the top ranking contestants had used the ensemble method, and particularly the xgboost [11].

Extreme gradient boosting (xgboost) is a regularized version of ordinary gradient boosting algorithm. The regularization term is introduced to penalize the complexity of the function, while making the result more robust to over fitting. The model learns an ensemble of boosted trees which makes precise adjustment between classification error and model complexity. Xgboost obtain an AMS score of 3.71885 on the private leaderboard [3]. Xgboost was the popular most in the competition due to its accuracy and efficiency.

Another most popular winning solution was proposed by Gabor Melis. Gabor's method was based on deep learning. Initially, a classifier was trained in order to predict the positive class probability. Then some of the test examples (those of the maximum predicted probabilities) were classified as positive. The cross-validation usually determine the decision threshold implemented on the training dataset to get the highest AMS score [10]. Additionally, this model is composed of an ensemble of 70 neural networks and the predicted likelihoods were combined as simple arithmetic averaging.

## 2. Proposed Higgs Deep Ensemble Classification (Higgs-DEC) Method

The proposed method includes, Deep Neural Network (DNN), Random forest and ensemble based classification techniques. In the ensemble based learning, "Deep Auto Encoder (DAE)" and "Random forest" are used as base classifiers and "Bagged Decision Tree" as a Meta-classifier. Using Meta-classifier, the predictions obtained from several base classifiers have been taken into account and a final decision is made on the classification. We weight the probability outputs of the base classifiers instead of predictions. $F(x) = \hat{p}$, where $\hat{p} = \{\hat{p}(y = 1|F, x), \hat{p}(y = 0|F, x)\}$. The response is simply the value with maximum probability.

$$\hat{y} = Arg\ max_{i \in \{1,0\}} \hat{p}(y = i|F, x)$$

Amongst all the classification techniques, the ensemble classifier turns to be the most superior classifier. In the proposed work, an ensemble based approach is used to solve this problem. As accuracy is not considered as a good measure for a highly unbalanced data, so area under the ROC cure is also reported. The online facility for performance calculation is to check the AMS score. The fully tuned Random forest and DAE gives the AMS score of 2.91 and 2.89 respectively. Moreover, the features from the second last layer of DAE stacked with the original features give improved results, and the AMS score of 3.42 is achieved. A block diagram of our proposed method is shown in Fig. 1.

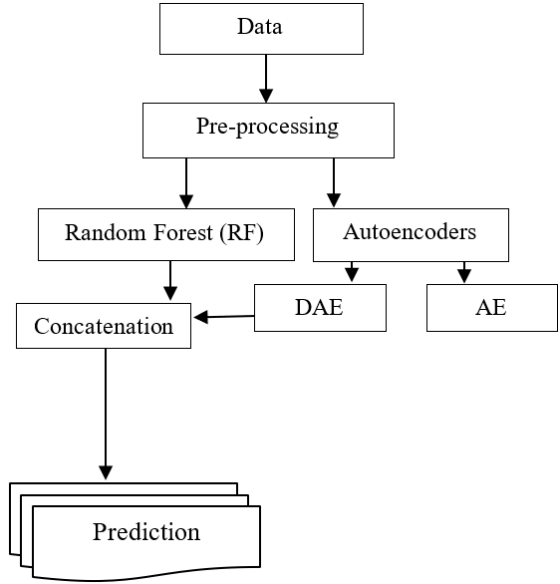

Figure 1: Basic block diagram of the proposed Higgs-DEC

## 3. Results

*3.1. Performance of Xgboost Classifier*

Xgboost is useful for many tasks especially for supervised learning. It provides parallel tree boosting which is very fast and accurate for many data science related problems [3], [12]. Xgboost execution speed is really fast as compared to other implementations.

The model parameters are selected such that, the objective loss function is minimum. In this case, the objective function comprises of the sum of training loss and regularization $Obj(\theta) = L(\theta) + \Omega(\theta)$. For a selected parameter, the objective function will be minimum, if both the training losses as well as the regularization of the model are minimized. The regularization term shows how well the model fits on the data, thus control the complexity of the model. Implementation of Xgboost [3] had performed very well on the Higgs boson data. We shall compare our results to it.

In the proposed work, Xgboost is implemented and accuracy of 87.25% is achieved on cross validation data. It is fast, efficient and has a parallel implementation of gradient boosting trees. It has tunable parameters, and the tuning of a parameter is simple. A small value of $\eta$ provides good learning results. The area under the ROC curve is an essential measure for the classifier. The performance of the classifier is high when AUROC is maximum.

### *3.2. Performance of Random Forest Classifier*

Random forest is a very easy machine learning algorithm. The ensemble is trained with "bagging" and "boosting" methods. In ensemble, the learning models increase the overall result. The information gained at each node is calculated. If it is maximum, then the classification performance is maximum.

In the proposed work, random forest is tuned with 500 trees. The model is trained with 70% of the training data and cross validated by the remaining 30% of the data. It gives the accuracy of 84% and the AUROC of 0.8643. Area under the ROC curve is responsible for high AMS score. The more the AUC, the more will be the AMS score. The selected model is then trained with total training data and is tested with total testing data. We get the AMS score of 2.915 from the online submission for the Random forest. From the confusion matrix, we can observe true positives, false positives, true negatives, false negatives, as well as the prediction rates. The accuracy of each class prediction is also visible from the confusion matrix.

### *3.3. Performance of Deep Neural Network*

Deep neural networks [13] have outperformed in some applications, and have proven to be a powerful tool in different areas [14]. Using DNN, the performance of different networks is studied in detail. The training data is very small as compared to testing data, highly imbalanced. The signals of Higgs are very little compared to the background signals [15].

The performance of networks from sparse to deep is evaluated in detail. The detection performance of the classifier increases for deep networks because the additional important features are extracted from the hidden layers.

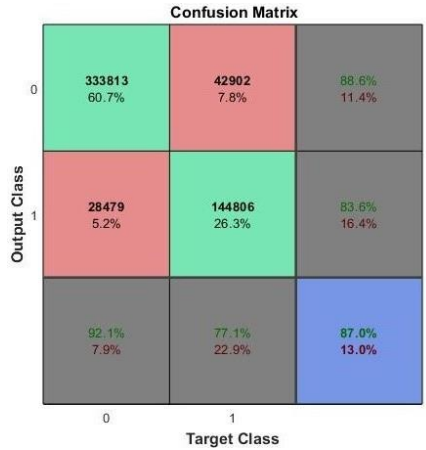

Figure 2: Confusion matrix of the proposed Higgs-DEC classifier

The autoencoder (AE) can extract the most important features, are expressive and scalable, and can predict the labels. Using DAE, stacked layers improve positive class prediction.

The performance of the DNN is evaluated in terms of AMS score, and is comparable to the ensemble method. The first layer of AE gives the accuracy of 80.9% and AUROC of 0.82, whereas three-layered network gives improved results. The accuracy is 83.6% and AUROC is 0.854. The AMS score for this network classifier is 2.89 for public leaderboard.

*3.4. Performance of the Proposed Higgs-DEC Classifier*

The ensemble method includes the results from different architectures of Random forest ensembles and DNNs. An architecture of 9 classifiers along with the improved feature space provides improved results. A good AMS score of 3.19 is achieved from this ensemble classifier. Another ensemble includes the predictions from 9 different Random forests and 6 different autoencoders. The ensemble can capture randomly the best result. The base classifiers are not to be too optimized, because they have to capture random points. The accuracy from this ensemble method is given below in the performance matrix. The results are the best of all the classifiers tried, the true class prediction is maximum as compared to all the classifiers tried on the data with an AUROC of 0.9 and AMS score of 3.42. The confusion matrix of the proposed Higgs-DEC is shown in Fig. 2.

| Classifier | Accuracy | AUROC | AMS Score |
|---|---|---|---|
| **RF** | 84 | 0.8643 | 2.915 |
| **AE** | 80.9 | 0.82 | 2.701 |
| **DAE** | 83.6 | 0.854 | 2.893 |
| **RF "A"** | 85.5 | 0.87 | 3.196 |
| **Ensemble** | 87.0 | 0.90 | 3.429 |

Table 1: Results comparison table

*3.5. Comparison and Analysis*

To compare the classifiers, online AMS score is calculated for the predictions. To get high AMS score, the true positive rate for a classifier must be high. The comparison includes the results from Random forest, single layer autoencoder, deep autoencoder, Random forest "A" and the ensemble method. In the Random forest "A", additional features from deep autoencoder are stacked with original feature space, and then predictions are taken from the Random forest. The performance in this case is slightly improved. In the ensemble method, deep autoencoders and Random forests are stacked as base classifiers. For the meta-classifier, the performance of many classifiers like linear SVM, Gaussian SVM, Bagged decision tree, Boosted decision tree and kernel SVM are checked. The best of all is found to be Bagged decision tree. The comparison of AMS score is shown in the Table 1. We measure the area under the curve (AUC) from the ROC as the accuracy is not a good measure for a highly unbalanced data. The online facility for performance calculation is to check the AMS score. The fully tune Random forest gives the accuracy of 84% and the AMS score of 2.91, whereas the DAE gives the accuracy of 83.6% and the AMS score of 2.89. Stacking of the features from the second last layer of DAE with the original features gives improved results, and the AMS achieved in this case was 3.19. The ensemble is ranked with the top AMS score of nearly 3.43, whereas the reported top AMS score is 3.81.

## 4. Conclusions

The real challenging aspect is that the nature of the Higgs boson signal is quite similar to the background signals. For a high AMS score, a low false-positive rate and a high true-positive rate is required. Accuracy has a minimal effect on the AMS score. The shallow architecture cannot capture the complicated and non-linear functions. The addition of layers to the neural network might improve the result. In this study, several individual classification systems are employed to differentiate the Higgs signal from the background noise. Additionally, we proposed an ensemble method based on random Forest and Deep Autoencoder, which performs better than all the individual classifiers by yielding an AMS score of 3.429 on the public leaderboard. In the future, we aim to employ the stacking of various Deep Neural Networks to improve the results further.